\documentclass[%
 aip,preprint,%
 amssymb, amsmath,%
 cha,%
]{revtex4-1}
\usepackage[T1]{fontenc}
\usepackage{array}
\usepackage{float}
\usepackage{multirow}
\usepackage{amsmath}
\usepackage{amssymb}
\usepackage{graphicx}
\usepackage{caption}
\usepackage{subcaption}
\usepackage{fancyref}
\usepackage{esint}
\draft 

\numberwithin{equation}{section}

\begin{document}


\title{Semi-analytic Shape Function of High-harmonic Electron Cyclotron Emission in Tenuous Plasma} 


\author{J. Leem}
\email[]{juneeokleem@gmail.com}
\author{J. Jo}
\affiliation{Department of Physics, Pohang University of Science and Technology, Pohang 37673, Republic of Korea}
\author{G. S. Yun}
\affiliation{Department of Physics, Pohang University of Science and Technology, Pohang 37673, Republic of Korea}
\affiliation{Division of Advanced Nuclear Engineering, Pohang University of Science and Technology, Pohang 37673, Republic of Korea}


\date{\today}

\begin{abstract}
Semi-analytic expressions for the electron cyclotron emission (ECE) shape function are developed for arbitrary high harmonics. The integrand of the ${m}^{\mathrm{th}}$ harmonic ECE shape function is fitted with the readily-integrable test function which is parameterized by plasma temperature $T_e$, harmonic number $m$ and emission angle $\theta $. Semi-analytic formulae for high harmonic ECE emissivity are obtained by integrating the test integrand with the fitting parameters gained from the regression analysis as well as the inductive studies for arbitrary cases. The developed expression matches the numerically-integrated ECE shape function very well and the overall differences between original shape functions and semi-analytic shape functions are evaluated. The expression can be used for rapid analysis of high-harmonic ECE spectra.
\end{abstract}

\pacs{empty}

\maketitle 


\section{Introduction}
Electron cyclotron emission (ECE) is an important phenomenon in a magnetized, high-temperature plasma in respect of its applicability in plasma diagnostics. ECE involves useful information such as plasma temperature, density, and energy distribution of electrons \cite{ref1}. More importantly, ECE frequency can be spatially localized when the plasma is subjected to the strong magnetic field \cite{ref2} so that the spectroscopic analyses of ECE provide the spatial profiles of plasma properties. In addition to this, the existence of ECE harmonics with their distinct characteristics enriches the plasma information significantly.

The fundamental and the ${2}^{\mathrm{nd}}$ harmonic of ECE are extensively used for electron temperature diagnostics owing to their optical thickness which enables the usage of black-body radiation \cite{ref3,ref4,ref5}. In contrast, the higher harmonics of ECE, whose optical depths are relatively thin, are potential candidates for the unusual diagnostics such as the measurements of non-Maxwelllian(energetic) electrons \cite{ref6,ref7,ref8}, and the magnetic fields near the magnetostar \cite{ref9}. For the diagnostic uses of ECE, the spectral interpretations of the emissivity in various plasma circumstances must be preceded in order to modify the spectra and their corresponding intensities into plasma parameters.

Emissivity $j\left(\omega \right)$ is expressed as a multiplication of two components: radiated power per unit solid angle from certain unit plasma volume ($\mathrm{\ }p=\int{j\left(\omega \right)d\omega }$), and a shape function (${\mathrm{\Phi }\left(\omega \right)=\ j\left(\omega \right)}/{p}$). The analyses of the ECE spectra require a huge number of corresponding shape functions for a given plasma circumstance in order to establish the emission profile \cite{ref10}. However, due to the massiveness of the computing processes as well as the complexities of the ECE shape function formulae, it is inefficient to adjust the numerical approaches on calculating ECE shape functions for a rapid interpretation of the ECE signal. For this reason, the detailed studies and the processes in developing the analytic expressions of the ECE shape functions will be introduced in the next section.\cite{ref11,ref12,ref13}

\section{SHAPE OF THE ELECTRON CYCLOTRON EMISSIVITY IN THE FREQUENCY DOMAIN}

In magnetized plasma, the electrons gyrate around the magnetic field lines and emit electromagnetic radiations at their fundamental cyclotron frequencies. The harmonic emissions can be generated when the trajectory of a single electron deviates from the circle, and their actual emission frequencies are broadened and shifted due to the relativistic Doppler effects. The resonance angular frequency ${\omega}_{m,\mathrm{res}}$ is determined as

\begin{equation}
{\omega}_{m,\mathrm{res}}=\frac{m{\omega}_c}{1-{\beta }_{\parallel }{\mathrm{cos} \theta \ }}\mathrm{\ \ \ \ }
\label{eq1}
\end{equation}
With $\beta={v_e}/{c}$, ${\omega}_c={eB_0}/{\gamma m_e}$, $\gamma={\left[1-\left({\beta}^2_{\bot}+{\beta}^2_{\parallel }\right)\right]}^{{-1}/{2}}$. The symbols $\parallel$ and $\bot$ represent the parallel and perpendicular components, and $\theta$ is the emission angle with respect to the external magnetic field.

In tenuous plasma, the inter-particle interactions are negligible so that the plasma emissivity, which is defined as the radiated power per unit volume, unit solid angle, and unit frequency, can be expressed as a sum of all the radiations from the individual electrons. Detailed derivations of the plasma emissivity for each harmonic radiation are listed in \cite{ref1}, and the expression of normalized spectral emissivity ($\int^{\infty}_0{\mathrm{\Phi }\left(\omega\right)d\omega}=1$, shape function) of ${2}^{\mathrm{nd}}$ harmonic ECE in local thermal equilibrium (Maxwell distributed electrons) is derived by S. K. Rathgeber \cite{ref10}.

\begin{equation}
\begin{split}
\mathrm{{\Phi}_2}\left(\omega;\theta,T_e,B_0\right) \ = 
& \frac{{\zeta }^{{7}/{2}}}{\sqrt{\pi }}\int^1_{-1}{d{\beta }_{\parallel }}\int^{\sqrt{1-{\beta }^2_{\parallel }}}_0{d{\beta }_{\bot }}{\beta }^5\\
& \times \delta \left(\left[1-{\beta }_{\parallel }{\mathrm{cos} \theta \ }\right]\omega -\frac{{\omega }_{2X}}{\gamma }\right) {\mathrm{exp} \left(-\zeta \left[{\beta }^2_{\bot }+{\beta }^2_{\parallel ~}\right]\right)\ }
\end{split}
\label{eq2}
\end{equation}
With $\zeta={m_{e}c^2}/{2T_e}$, ${\omega }_{2X}=2{eB_0}/{m_e}$. General expression of ${m}^{\mathrm{th}}$ harmonic ECE shape function can be obtained by implementing Rathgeber's idea \cite{ref10} into ${m}^{\mathrm{th}}$ harmonic emissivity \cite{ref1}
\begin{equation}
\begin{split}
{\mathrm{\Phi}}_m\left(\omega;\theta,T_e,B_0\right) \ =
& \frac{2{\zeta }^{(2m+3)}}{\sqrt{\pi }m!}\int^1_{-1}{d{\beta }_{\parallel }}\int^{\sqrt{1-{\beta }^2_{\parallel }}}_0{d{\beta }_{\bot }}{\beta }^{2m+1} \\
& \times \delta \left(\left[1-{\beta }_{\parallel }{\mathrm{cos} \theta}\right]\omega -\frac{{\omega }_m}{\gamma }\right)
{\mathrm{exp} \left(-\zeta \left[{\beta }^2_{\bot }+{\beta }^2_{\parallel ~}\right]\right)\ }
\end{split}
\label{eq3}
\end{equation}
Where ${\omega}_m={meB_0}/{m_e}$ is the cold resonance frequency of ${m}^{\mathrm{th}}$ harmonic ECE. In order to integrate the delta function in \eqref{eq3}, following identity can be used.
\begin{equation}
\int^b_a{f(x)\delta(h\left(x\right))dx}=\sum_i{\frac{f(x_i)}{\left|h^{\prime}(x_i)\right|}}
\label{eq4}
\end{equation}
Where $x_i$ are roots of $h\left(x\right)$ in $\left[a,b\right]$. Regarding $\gamma$ = ${\left[1-\left({\beta}^2_{\bot}+{\beta}^2_{\parallel}\right)\right]}^{{-1}/{2}}$, ${\beta}^2_{\bot}=1-{\beta}^2_{\parallel}-\left({{\omega }^2}/{{\omega }^2_m}\right){\left(1-{\beta }_{\parallel}{\mathrm{cos}\theta}\right)}^2$ is the only root of the argument of the delta function in \eqref{eq3} within the integration range. Adjusting \eqref{eq4} to \eqref{eq3} with $\mu={{\omega}^2}/{{\omega}^2_m}$.
\begin{equation}
\begin{split}
{\mathrm{\Phi }}_m\left(\omega;\theta,T_e,B_0\right) \ = 
& \frac{2{\zeta }^{{(2m+3)}/{2}}\omega }{\sqrt{\pi}m!{\omega }^2_m}\int^1_{-1}{d{\beta }_{\parallel }} \left(1-{\beta }_{\parallel }{\mathrm{cos} \theta}\right) \\
& \times {\left[1-{\beta }^2_{\parallel ~}-\mu {\left(1-{\beta }_{\parallel }{\mathrm{cos} \theta}\right)}^2\right]}^m 
{\mathrm{exp}\left[-\zeta \{1-\mu{\left(1-{\beta }_{\parallel }{\mathrm{cos}\theta}\right)}^2\}\right]}
\end{split}
\label{eq5}
\end{equation}
Because ${\beta}_{\bot}$ is a real, ${\beta}_{\bot}$ = $1-{\beta}^2_{\parallel} - \mu{\left(1-{\beta}_{\parallel}{\mathrm{cos} \theta}\right)}^2$ $\ge$ $0$. Thus, the following relation is satisfied with $\eta =1+\mu {{\mathrm{cos}}^2 \theta \ }$
\begin{equation}
\begin{split}
-1 \le \frac{\mu {\mathrm{cos} \theta \ }-\sqrt{1-\mu {{\mathrm{sin}}^2 \theta \ }\ }}{\eta } \le {\beta }_{\parallel }
\le \frac{\mu {\mathrm{cos} \theta \ }+\sqrt{1-\mu {{\mathrm{sin}}^2 \theta \ }\ }}{\eta } \le 1\mathrm{\ \ \ }
\end{split}
\label{eq6}
\end{equation}
The integration range of \eqref{eq5} is modified as
\begin{equation}
\begin{split}
{\mathrm{\Phi }}_m\left(\omega;\theta,T_e,B_0\right) \ = & \frac{2{\zeta }^{{(2m+3)}/{2}}\omega }{\sqrt{\pi}m!{\omega }^2_m}\int^{{\beta}_{\parallel +}}_{{\beta }_{\parallel -}}{d{\beta }_{\parallel }} \left(1-{\beta }_{\parallel }{\mathrm{cos} \theta}\right) \\
& \times {\left[1-{\beta }^2_{\parallel ~}-\mu {\left(1-{\beta }_{\parallel }{\mathrm{cos} \theta}\right)}^2\right]}^m 
{\mathrm{exp}\left[-\zeta \{1-\mu{\left(1-{\beta }_{\parallel }{\mathrm{cos}\theta}\right)}^2\}\right]}
\end{split}
\label{eq7}
\end{equation}
Where ${\beta }_{\parallel \pm }=\ {\left(\mu {\mathrm{cos} \theta \ }\pm \sqrt{1-\mu {{\mathrm{sin}}^2 \theta \ }\ }\right)}/{\eta }$. \eqref{eq7} can be simplified by substituting $\alpha ={\left(1-{\beta }_{\parallel }{\mathrm{cos} \theta \ }\right)}^2$
\begin{equation}
\begin{split}
{\mathrm{\Phi}}_m\left(\omega;\theta,T_e,B_0\right) \ = 
& \frac{{\zeta }^{{(2m+3)}/{2}}\omega }{\sqrt{\pi }m!{\omega }^2_m{\mathrm{cos} \theta \ }}\int^{{\alpha }_+}_{{\alpha }_-}{~d\alpha} \\
& \times {\left[1-{\left(\frac{1-\sqrt{\alpha }}{{\mathrm{cos} \theta \ }}\right)}^2-\mu \alpha \right]}^m {\mathrm{exp} \left[-\zeta \left(1-\mu \alpha \right)\right]}
\end{split}
\label{eq8}
\end{equation}
Where 
\begin{equation}
{\alpha }_{\pm }={\left(\frac{1\pm {\mathrm{cos} \theta \ }\sqrt{1-\mu {{\mathrm{sin}}^2 \theta \ }}}{\eta }\right)}^2
\label{eq9}
\end{equation}

One may integrate \eqref{eq8} \textbf{$m$} times repeatedly by parts in order to get the analytic expression of ${m}^{\mathrm{th}}$ harmonic ECE shape function. According to the actual work on ${2}^{\mathrm{nd}}$ X-mode ECE shape function \cite{ref10}, integration by parts becomes increasingly difficult and impractical for higher harmonics though it may not be impossible. Therefore, we studied and developed more practical approach to estimate the shape of the cyclotron emission line including high harmonic emissions.

\section{SEMI-ANALYTIC EXPRESSEION OF ECE SHAPE FUNCTION}

One of the basic principles of definite integration is that it yields the area under integrand in Cartesian coordinates. Therefore, the practical approach starts with finding the readily integrable test function which fits into the integrand in the integration range. The integrand $g\left(\alpha \right)$ of the generalized shape function in \eqref{eq8} is

\begin{equation}
g\left(\alpha\right)={\left[1-{\left(\frac{1-\sqrt{\alpha}}{{\mathrm{cos}\theta}}\right)}^2-\mu\alpha\right]}^m \mathrm{exp} \left[-\zeta\left(1-\mu\alpha\right)\right]
\label{eq10}
\end{equation}

\begin{figure}[!]
	\includegraphics[width=0.30\textwidth]{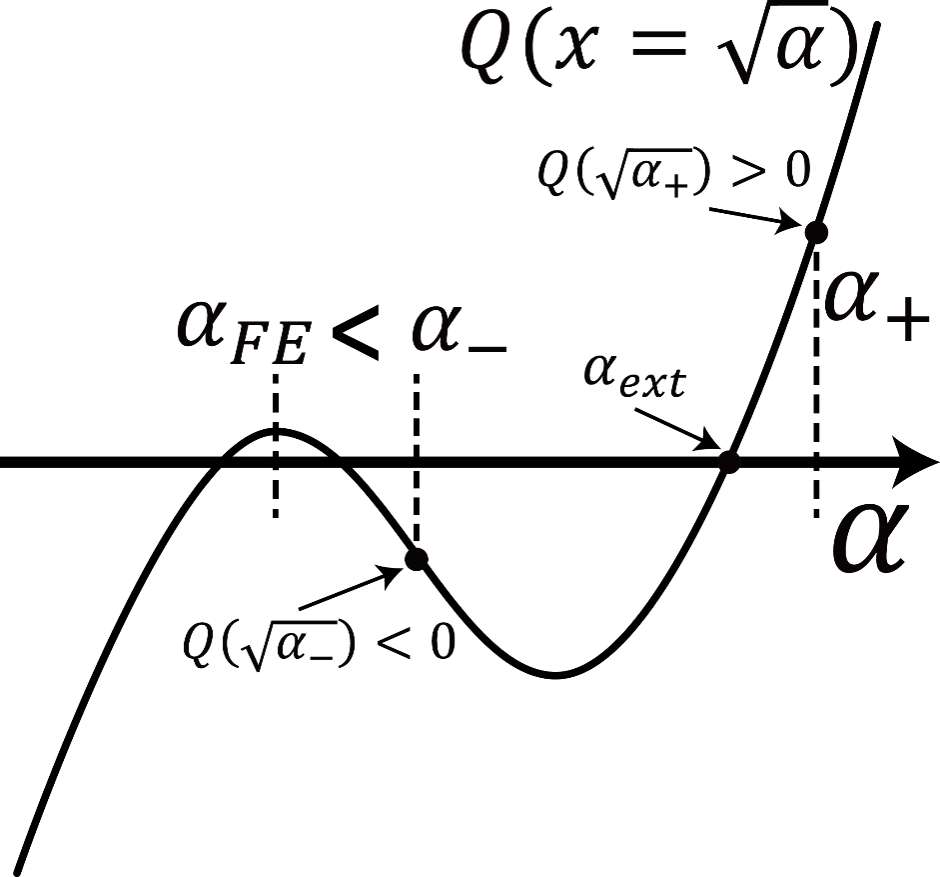}
	\caption{Overall shape of $Q(x)$ in $\theta\sim90^{\circ}$ case. $\alpha_{\pm}$ is in \eqref{eq9}, and $\alpha_{\mathrm{FE}}$ is the first extremum point. $Q(x)$ has three real root in $\alpha$ domain.}
	\label{fig1}
\end{figure}

The test function can be inferred from the graphical tendency as well as the inflection and the extremums of the original function. In the low $\alpha $ regime, the expression in square bracket dominates the integrand, and exponential function dominates the integrand in the high $\alpha $ regime. Consequently the test function $f\left(\alpha \right)$ for the \eqref{eq10} is determined as follow.

\begin{equation}
f\left(\alpha \right)=A{\left(\alpha -{\alpha }_+\right)}^k{\mathrm{exp} \left(-B{\left(\alpha -{\alpha }_+\right)}^2\right)\ }
\label{eq11}
\end{equation}
Parameters $A$ and $B$ are derived by matching the extremums of $f\left(\alpha \right)$ and $g\left(\alpha \right)$.
\begin{equation}
\begin{split}
& A=\frac{g_{\mathrm{ext}}}{{\left({\alpha }_{\mathrm{ext}}-{\alpha }_+\right)}^k{\mathrm{exp} \left(-B{\left({\alpha }_{\mathrm{ext}}-{\alpha }_+\right)}^2\right)\ }}\\
& B=\frac{k}{2{\left({\alpha }_{\mathrm{ext}}-{\alpha }_+\right)}^2}
\end{split}
\label{eq12}
\end{equation}
In this case, $\left({\alpha }_{\mathrm{ext}}, g_{\mathrm{ext}}\right)$ is assumed to be the extremum of $g\left(\alpha \right)$. Because $g^{\prime}\left({\alpha }_{\mathrm{ext}}\right)$ is zero by definition, ${\alpha }_{\mathrm{ext}}$ is calculated by solving the following equation.
\begin{equation}
\begin{split}
\left(\mu+\frac{1}{{{\mathrm{cos}}^2\theta}}\right){{\alpha}_{ext}}^{{3}/{2}}-\left(\frac{2}{{{\mathrm{cos}}^2 \theta}}\right){{\alpha }_{ext}}
+ \left(\frac{m}{\zeta }+\frac{{m}/{\zeta \mu }+1}{{{\mathrm{cos}}^2 \theta \ }}-1\right){{\alpha }_{ext~}}^{{1}/{2}} = \frac{{m}/{\zeta \mu }}{{{\mathrm{cos}}^2 \theta \ }}
\end{split}
\label{eq13}
\end{equation}
Which is the form of the cubic polynomial \cite{ref11}.
\begin{equation}
Q\left(x\right)=\eta x^3-2x^2+\left(m\epsilon\eta+{{\mathrm{sin}}^2\theta}\right) x-m\epsilon =0
\label{eq14}
\end{equation}
Where $x=\sqrt{\alpha }$, $\epsilon =1/\zeta \mu $, $\eta ={\mathrm{1+}\mu \mathrm{cos}}^2\theta $. The cubic polynomial $Q\left(x\right)$ has only one real root in $\left[{\alpha }_-,{\alpha }_+\right]$ and therefore the root is the biggest among the roots of $Q\left(x\right)$ (FIG. \ref{fig1}). Hence, the cubic polynomial $Q\left(x\right)$ follows the hyperbolic solution for one real root in \cite{ref12}

\begin{equation}
\begin{split}
& p=\frac{3\eta \left({{\mathrm{sin}}^2 \theta +m\epsilon \eta \ }\right)-4}{3{\eta }^2}\\
& q=\frac{9\eta \left(2{{\mathrm{sin}}^2 \theta}-m\epsilon \eta \right)-16}{27{\eta }^3}
\end{split}
\label{eq15}
\end{equation}

\begin{equation}
\begin{split}
&if \ \ \ \ p< 0 \ , \ \ \
{\alpha }_{\mathrm{ext}}={\left\{2\sqrt{-\frac{p}{3}}{{\mathrm{cos}\left[\frac{1}{3}{{{\mathrm{cos}}^{-1} \left(\frac{3q}{2p}\sqrt{-\frac{3}{p}}\right)}}\right]}}+\frac{2}{3\eta}\right\}}^2\\
\\
&else \ \ \ p \ge 0 \ , \ \ \
{\alpha }_{\mathrm{ext}}={\left\{-2\sqrt{\frac{p}{3}}{{\mathrm{sinh}\left[\frac{1}{3}{{\mathrm{sinh}}^{-1} \left(\frac{3q}{2p}\sqrt{\frac{3}{p}}\right)}\right]}}+\frac{2}{3\eta }\right\}}^2
\end{split}
\label{eq16}
\end{equation}

\begin{figure}[!]
	\includegraphics[width=0.98\textwidth]{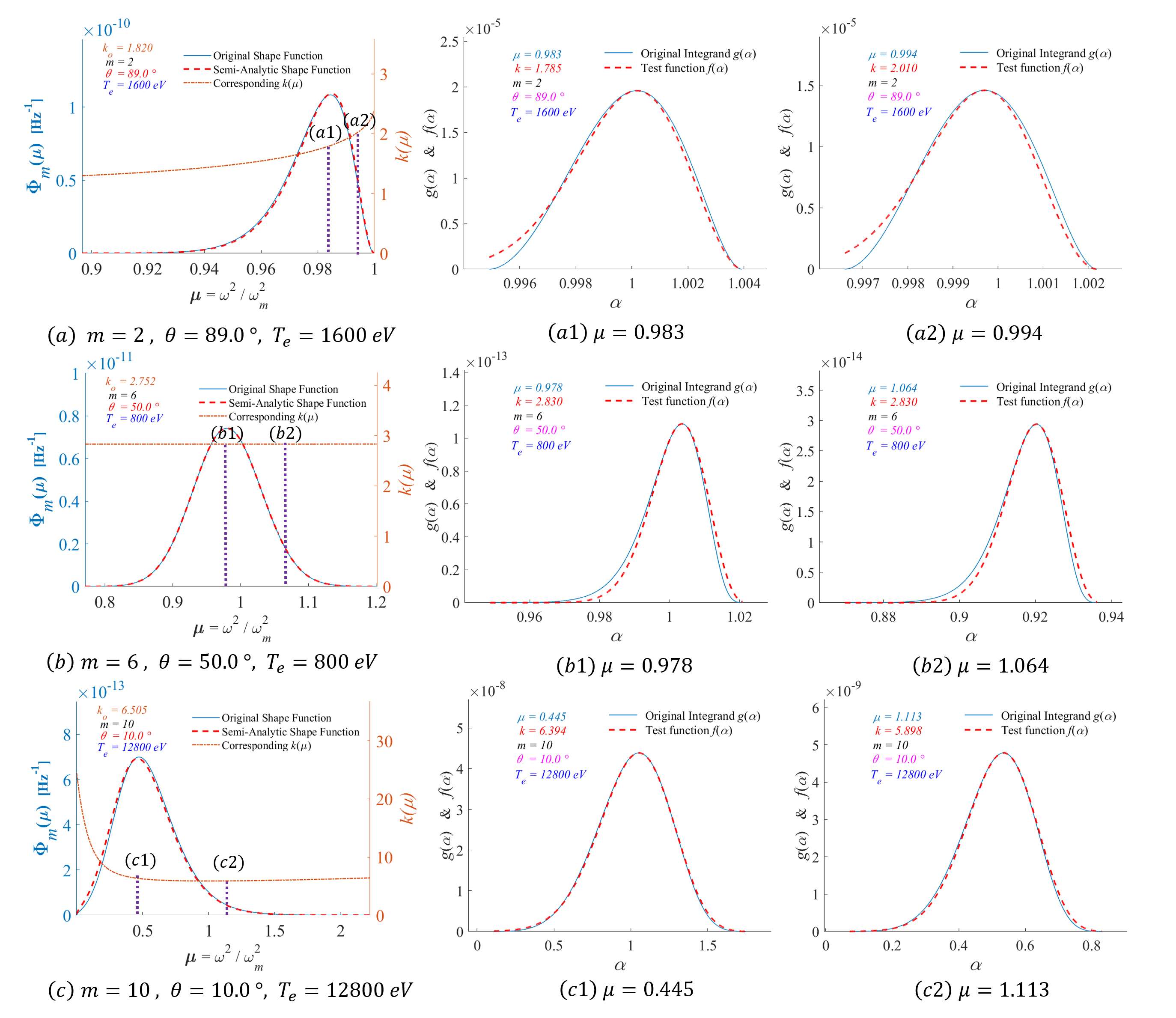}
	\caption{Original shape functions and their corresponding test functions are plotted by using the representative parameter $k_o$ which is obtained from the regression analysis. Subplot show how the test function $f\left(\alpha \right)$ is fitted to the original integrand $g\left(\alpha \right)$ for a certain $\mu$. In most cases, $k$ is almost constant for whole effective $\mu$ range as in (b), but (a) and (c) show two exceptional cases.}
	\label{fig2}
\end{figure}

Again, the main goal of this work is to establish the test function $f\left(\alpha \right)$ as a function of independent variables $\left(\omega;m,\theta,T_e,B_0\right)$. Once the evaluation of $\left({\alpha}_{\mathrm{ext}}, g_{\mathrm{ext}}\right)$ is done, the remaining task is to estimate $k$ in \eqref{eq11} \& \eqref{eq12}. The Trust-Region algorithm (also known as restricted-step method \cite{ref13}) is taken as a regression analysis tool to determine the parameter $k$ which is optimized to give the minimum residuals between the test function $f\left(\alpha \right)$ and the original integrand $g\left(\alpha \right)$ in the integration range $\left[{\alpha }_-,{\alpha }_+\right]$ (FIG. \ref{fig2}). This implies that for every normalized frequencies $\mu $ (FIG. \ref{fig2})  of the effective emissivity range (FIG. \ref{fig5b}), there exists corresponding parameters $k(\mu )$ such that in order to obtain the shape function with high frequency resolution, the semi-analytic approach consequently converges to the numerical analysis.

\begin{figure}[!]
\centering
	\includegraphics[width=0.75\textwidth]{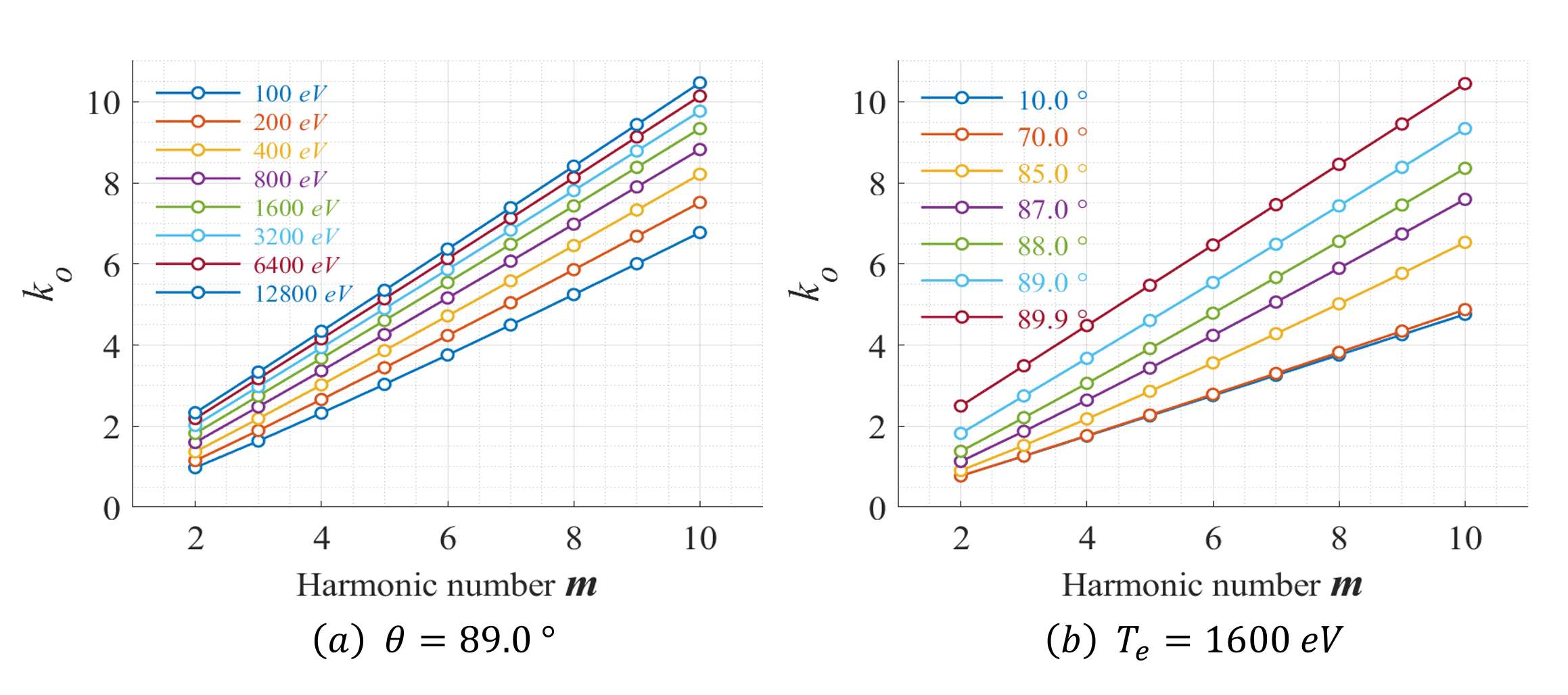}
	\caption{Linear correlation between harmonic number $m$ and $k_o$. This implies that the $k_o$ can be replaced by $d_o=m/k_o$}
	\label{fig3}
\end{figure}

\begin{figure}[!]
	\includegraphics[width=0.95\textwidth]{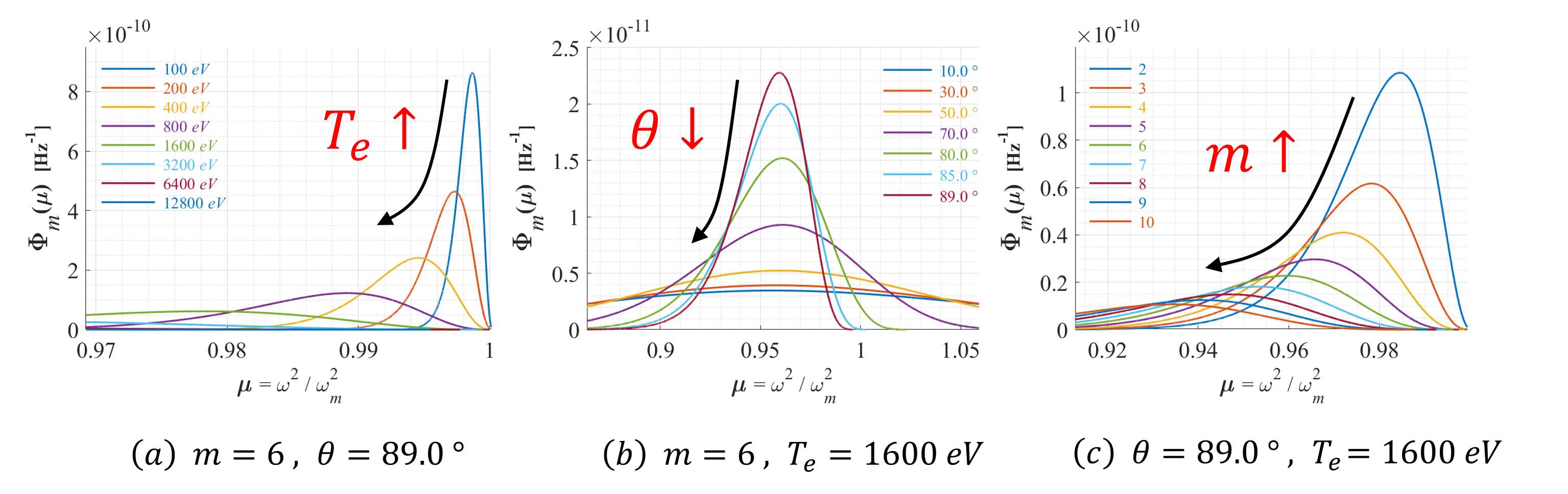}
	\caption{Individual effects of plasma temperature $T_e$, harmonic number $m$ and emission angle $\theta$ on the shape function.}
	\label{fig4}
\end{figure}

\begin{figure}[!]
	\includegraphics[width=0.95\textwidth]{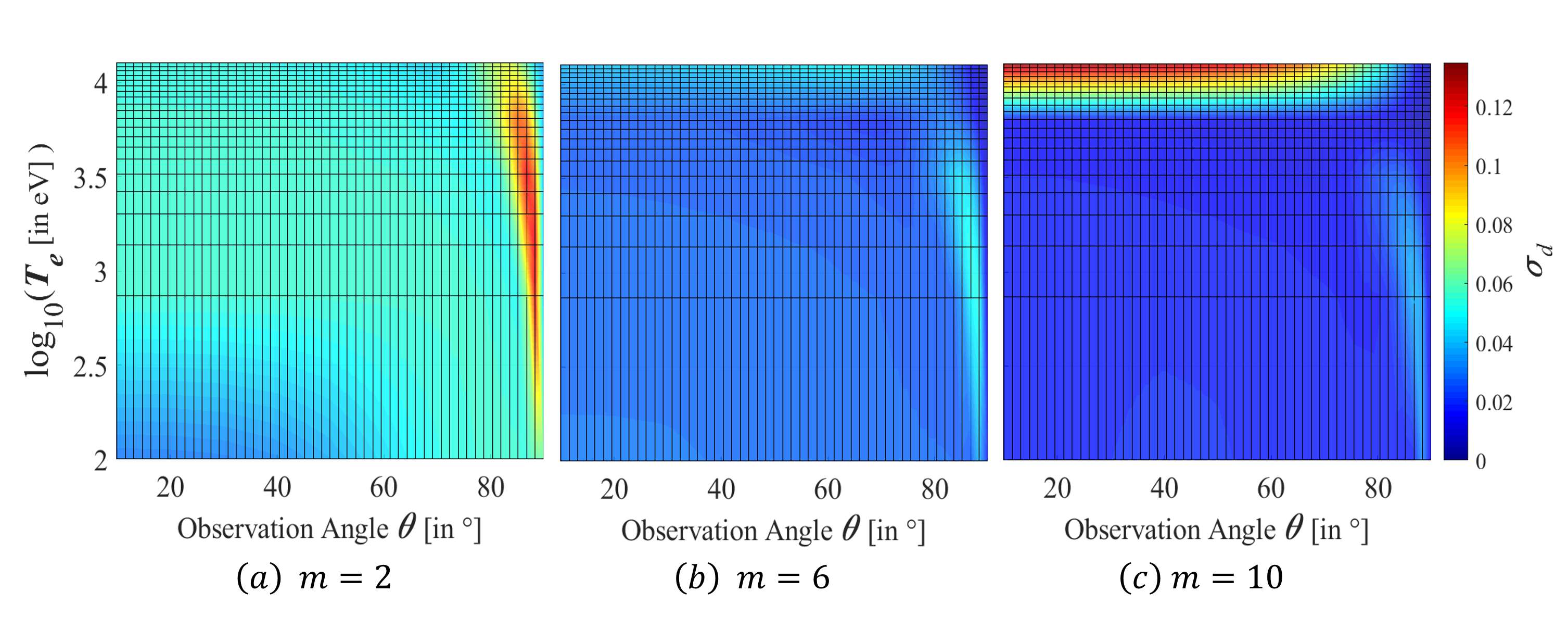}
	\caption{Plasma temperature $T_e$, harmonic number $m$ and emission angle $\theta$ effects on the $\mathrm{Deviation} \  {\sigma }_d$. Overall deviations do not exceed \textbf{0.15} even in the worst cases.This results inductively certificate the usage of $d_o\left(m,\theta,T_e\right)$}
	\label{fig5}
\end{figure}

\begin{figure}[!]
	\includegraphics[width=0.95\textwidth]{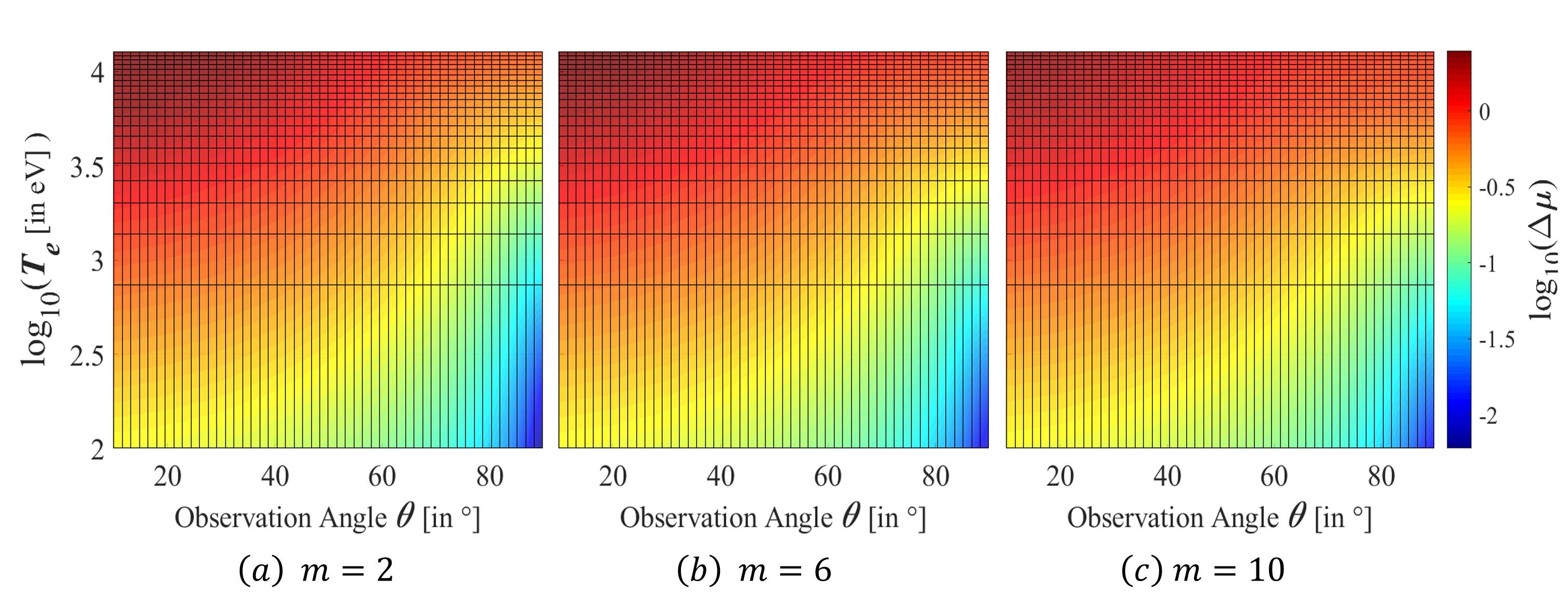}
	\caption{Doppler broadening effects of the shape function at a glance. Effective emissivity range $\Delta \mu$ is broadened as $\theta$ goes $0^{\circ}$ and $T_e$ increases.}
	\label{fig5b}
\end{figure}

Regarding the linear dependence of the parameter $k$ on the harmonic number $m$ (FIG. \ref{fig3}), the linear correlation coefficient $d=m/k$ now replaces the role of the fitting parameter $k$. Instead of calculating every single corresponding parameters $d(\mu )=m/k(\mu )$ , the representative parameter $''d_o=m/k_o''$ is adopted and will be used in the entire $\mu$ range to retain our approach not numerical but semi-analytic. Here, individual effects of plasma temperature $T_e$, harmonic number $m$ and emission angle $\theta$ on the shape function are independently investigated (FIG. \ref{fig4}) to support that implementation of the representative parameter $d_o=m/k_o$ is inductively logical. Overall deviations of $d\left(\mu;m,\theta,T_e\right)$ from $d_o\left(m,\theta,T_e\right)$ are evaluated through calculating ${\sigma }_d$ in \eqref{eq17}. Here, ${\sigma }_d$ is weighted and normalized by the original shape function ${\mathrm{\Phi }}_{m,\mathrm{Ori}}$ to ensure its representativeness in the whole $\mu$ range. (FIG. \ref{fig5})

\begin{equation}
\begin{split}
\mathrm{Deviation} \ {\sigma }_d\left(m,\theta ,T_e\right) =  \frac{\int{~d\mu \ {\mathrm{\Phi }}_{m,\mathrm{Ori}}\left(\mu;\theta,T_e\right)}\left|d_o\left(m,\theta,T_e\right)-d\left(\mu;m,\theta,T_e\right)\right|}{\int{~d\mu \  {\mathrm{\Phi}}_{m,\mathrm{Ori}}\left(\mu;\theta,T_e\right)}\left|d_o\left(m,\theta,T_e\right)\right|} 
\end{split}
\label{eq17}
\end{equation}

As a result, the reliability of $d_o\left(m,\theta,T_e\right)$ is proved except some marginal cases such as that the plasma temperature is over 5 $\mathrm{keV}$ and the harmonic number is sufficiently large. Now, the semi-analytic formula of the shape function can be expressed as
\begin{equation}
\begin{split}
{\mathrm{\Phi}}_m \left(\omega;\theta,T_e,B_0\right)=\frac{{\zeta }^{{(2m+3)}/{2}}\omega }{\sqrt{\pi }m!{\omega }^2_m{\mathrm{cos} \theta}} 
\int^{{\alpha }_+}_{{\alpha }_-}{~d\alpha \ A{\left(\alpha -{\alpha }_+\right)}^{{m}/{d_o}}{\mathrm{exp} \left(-B{\left(\alpha -{\alpha }_+\right)}^2\right)\ }}
\end{split}
\label{eq18}
\end{equation}
Substituting $x=\ B{\left(\alpha -{\alpha }_+\right)}^2$ \& $x_{\pm }=\ B{\left({\alpha }_{\pm }-{\alpha }_+\right)}^2$,
\begin{equation}
\begin{split}
{\mathrm{\Phi }}_m \left(\omega;\theta,T_e,B_0\right)=\frac{{\zeta }^{{(2m+3)}/{2}}\omega }{\sqrt{\pi }m!{\omega }^2_m{\mathrm{cos} \theta \ }} 
\int^0_{x_-}{dx \ \frac{1}{2}AB^{-{(m+d_o)}/{2 d_o}} x^{{(m-d_o)}/{2d_o}}e^{-x}}
\end{split}
\label{eq19}
\end{equation}
Note that the integrand of \eqref{eq18} converges to zero as ${\alpha \to \alpha }_-$, thus the integration range$\left[x_-,0\right]$ in \eqref{eq19} can be replaced by $\left[\infty,0\right]$ as long as $x=k{{\left(\alpha -{\alpha}_+\right)}^2}/{2{\left({\alpha }_{\mathrm{ext}}-{\alpha }_+\right)}^2}\ >0$. By the definition of Gamma function $\mathrm{\Gamma }\left(z\right)=\ \int^{\infty }_0{x^{z-1}e^{-x}dx}$
\begin{equation}
\begin{split}
{\mathrm{\Phi }}_m \left(\omega;\theta,T_e,B_0\right)=&\frac{{\zeta }^{{(2m+3)}/{2}}\omega }{\sqrt{\pi }m!{\omega }^2_m{\mathrm{cos} \theta \ }}\left|\frac{1}{2}AB^{-{(m+d_o)}/{2 d_o}} \mathrm{\Gamma}\left(\frac{m+d_0}{2 d_o}\right) \right|
\end{split}
\label{eq20}
\end{equation}
Therefore, the final expression of semi-analytic ${m}^{\mathrm{th}}$ harmonic ECE shape function becomes
\begin{equation}
\begin{split}
{\mathrm{\Phi}}_m \left(\omega;\theta,T_e,B_0\right)=&\frac{{\zeta }^{{(2m+3)}/{2}}\omega }{\sqrt{\pi }m!{\omega }^2_m{\mathrm{cos} \theta \ }}{\left(\frac{2d_o}{m}\right)}^{{(m+d_o)}/{2 d_o}} \\
& \times 
\left|g_{\mathrm{ext}}\left({\alpha }_{\mathrm{ext}}-{\alpha}_+\right){\mathrm{exp}\left(\frac{m}{2d_o}\right)}\mathrm{\Gamma}\left(\frac{m+d_0}{2 d_o}\right)\right|
\end{split}
\label{eq21}
\end{equation}

\section{DETERMINING THE PARAMETERS FOR SEMI-ANALYTIC EXPRESSION}

Although the semi-analytic expression for the ${m}^{\mathrm{th}}$ harmonic ECE shape function is established in \eqref{eq21}, it is still impossible to get the actual values of normalized emissivity without knowing the representative parameter $d_o=m/k_o$. Moreover, even if the representative parameter for a certain plasma and observation condition is determined through the regression analysis, it does not provide the prospect of further information in various circumstances. Consequentially, more comprehensive and inductive studies are significantly important to determine the corresponding representative parameters for the arbitrary cases.

\begin{figure}[!]
	\includegraphics[width=0.95\textwidth]{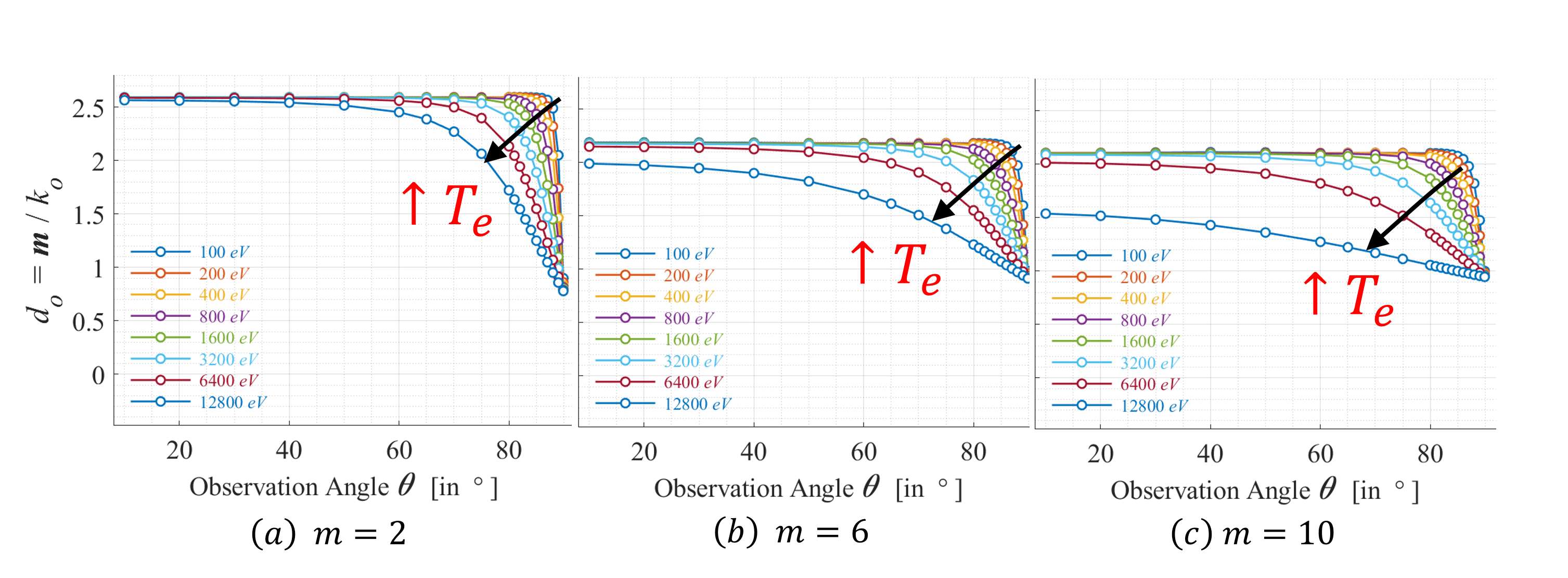}
	\caption{$T_e$ denpendence of $d_o$ in various $m$. This correlation has a similarity to Fermi-Dirac distribution.}
	\label{fig6}
\end{figure}

 In order to get the insight on how the corresponding representative parameter $d_o$ varies with $(m,\theta,T_e)$, a number of regression analyses are performed between the original ECE shape functions which are numerically calculated and semi-analytic expressions which are calculated by \eqref{eq21}. The evaluation ranges of $(m,\theta ,T_e)$ are listed in (TABLE. 1). If the plasma temperature is less than 5 keV, a significant correlation can be found among $(d_o,\theta,T_e)$ (FIG. \ref{fig6}). Thus, the correlation equation can be derived as \eqref{eq22} based on the graphical similarity to Fermi-Dirac distribution.

\begin{equation}
d_o\left(m;\theta,T_e\right)=\frac{\alpha }{1+\ \beta \left\{{\gamma \ {T_e}^{-1} \left(\theta - {\pi }/{2}\right)}\right\}}
\label{eq22}
\end{equation}

\begin{figure}[!]
	\includegraphics[width=0.95\textwidth]{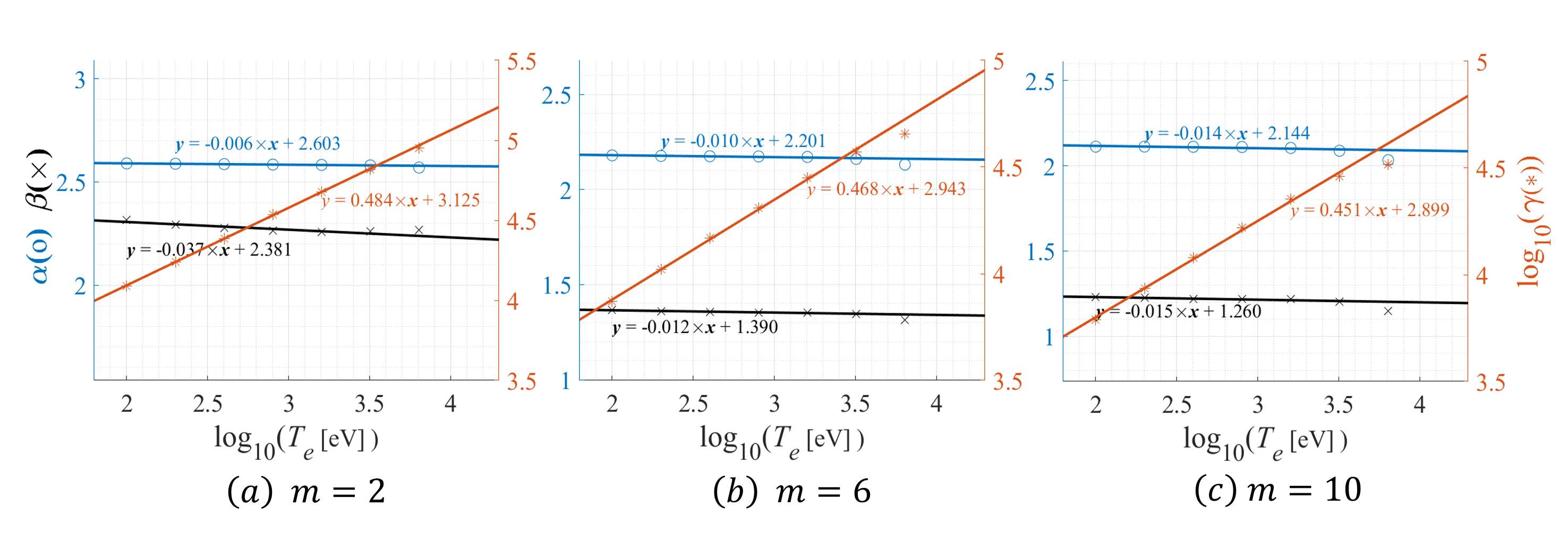}
	\caption{Individual coefficients in \eqref{eq22} are obtained through the regression analysis with linear fitting.}
	\label{fig7}
\end{figure}

The coefficients $\alpha, \beta, \gamma $ are determined from the secondary regression analysis in ${{\mathrm{log} \ T_e}}$ domain (FIG. \ref{fig7}). However, there still exists small $m$ dependences on the coefficients. The rough numbers of $\alpha, \beta, \gamma$ with different $m$ are listed in (TABLE. 2), therefore the final expression of the $d_o\left(m, \ \theta[\mathrm{rad.}], \ T_e[\mathrm{eV}]\right)$ becomes,

\begin{equation}
\begin{split}
d_o\left(m,\theta,T_e\right) = 
\frac{{\alpha }_2+\ {\alpha }_1{{\mathrm{log}}_{10} \left(T_e\right)\ }}{1+\ \left\{{\beta }_2+\ {\beta}_1{{\mathrm{log}}_{10} \left(T_e\right)\ }\right\}\left\{{\gamma }_2{T_e}^{{(\gamma }_1-1)}\left(\theta -{\pi }/{2}\right)\right\}}
\end{split}
\label{eq23}
\end{equation}

\begin{figure}[!]
	\includegraphics[width=0.95\textwidth]{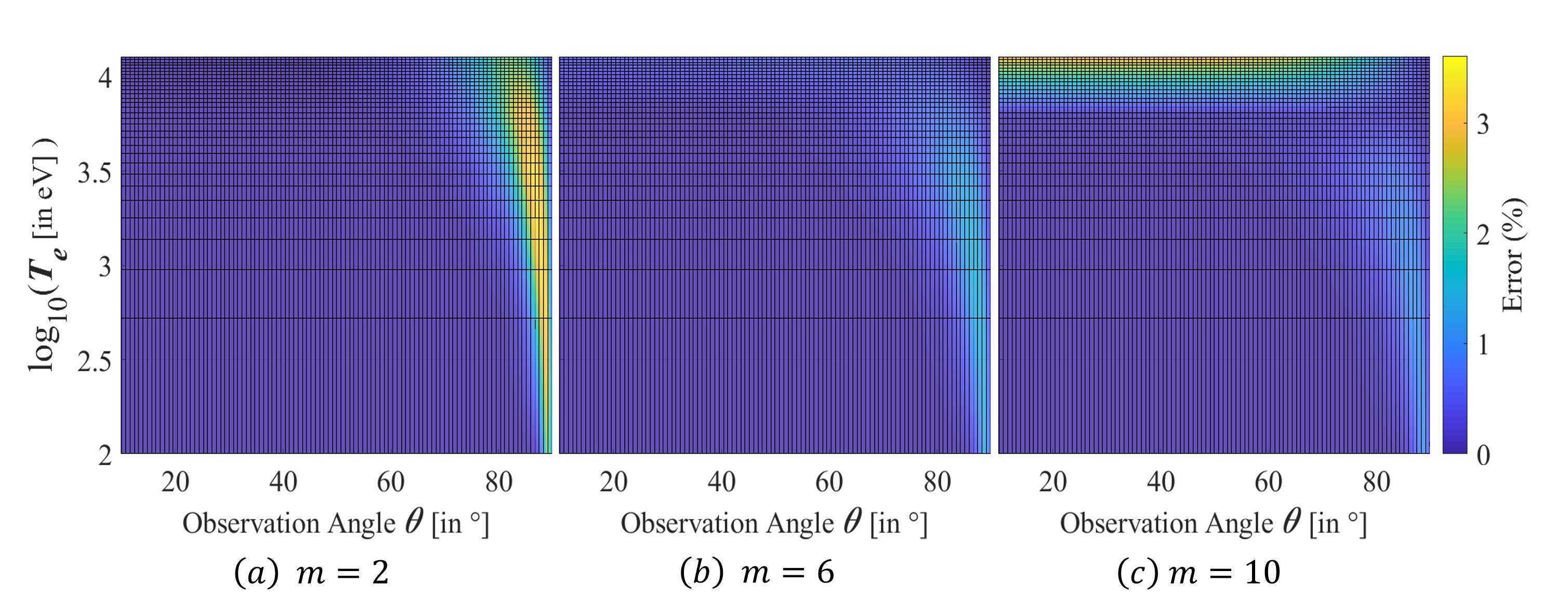}
	\caption{Overall difference [$\mathrm{Error} \ (\%)$] between the original shape function and the semi-analytic expression is no larger than 4$\%$. There is a close resemblance to $\mathrm{Deviation} \ {\sigma }_d$ in (FIG. \ref{fig5}) }
	\label{fig8}
\end{figure}

Finally, the differences between the original ECE shape functions and the semi-analytic results are evaluated using \eqref{eq24} which adds up the whole differences in effective emissivity range of $\mu$ (FIG. \ref{fig8}).

\begin{equation}
\begin{split}
&\mathrm{Error} \ \left(\%\right) = 
\frac{\int{~d\mu \ \left|{\mathrm{\Phi }}_{m,\mathrm{Ori}}\left(\mu;\theta,T_e\right)-{\mathrm{\Phi }}_{m,Test}\left(\mu;\theta,T_e\right)\right|}}{\int{~d\mu \ {\mathrm{\Phi}}_{m,\mathrm{Ori}}\left(\mu;\theta,T_e\right)}}
\end{split}
\label{eq24}
\end{equation}

\section{Conclusion}

The semi-analytic expression of ${m}^{\mathrm{th}}$ harmonic ECE shape function has been established with readily integrable function. The fitting parameters of the integrable function are examined in various plasma circumstances, and their resultant differences between the semi-analytic expressions and the numerically calculated ECE shape functions have been evaluated to estimate the applicable boundaries of this work, in particular, on fusion plasma regime. This work can be implemented to high-speed analysis of ECE diagnostics as well as their corresponding synthetic diagnostics \cite{ref14,} which are usually utilized for the verification and validation tools.


%
%

%

\begin{acknowledgments}
This work was supported by the National Research Foundation of Korea under grant \textbf{No. NRF-2017M1A7A1A03064231} and \textbf{BK21+} program.
\end{acknowledgments}


%

\pagebreak

\pagebreak

\begin{table}[!]
	\includegraphics[width=0.8\textwidth]{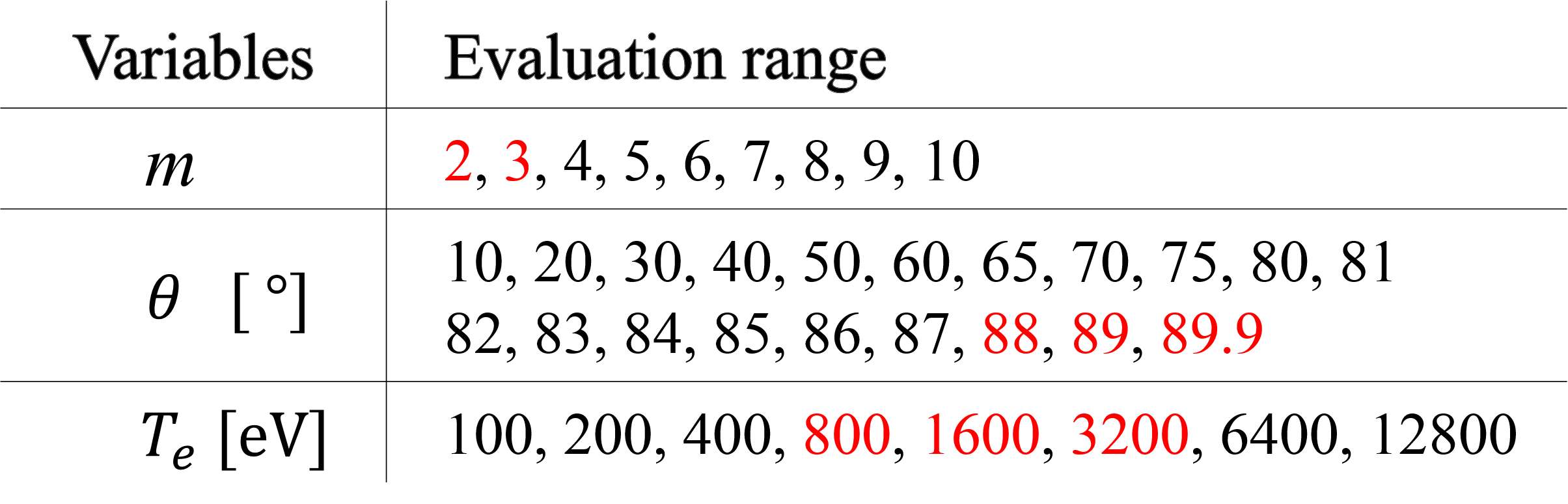}
	\caption{The evaluation ranges of $(m,\theta,T_e)$ for the further analysis in this section. Common values in the fusion plasma circumstance are emphasized in red.}
	\label{table1}
\end{table}

\begin{table}[!]
	\includegraphics[width=0.8\textwidth]{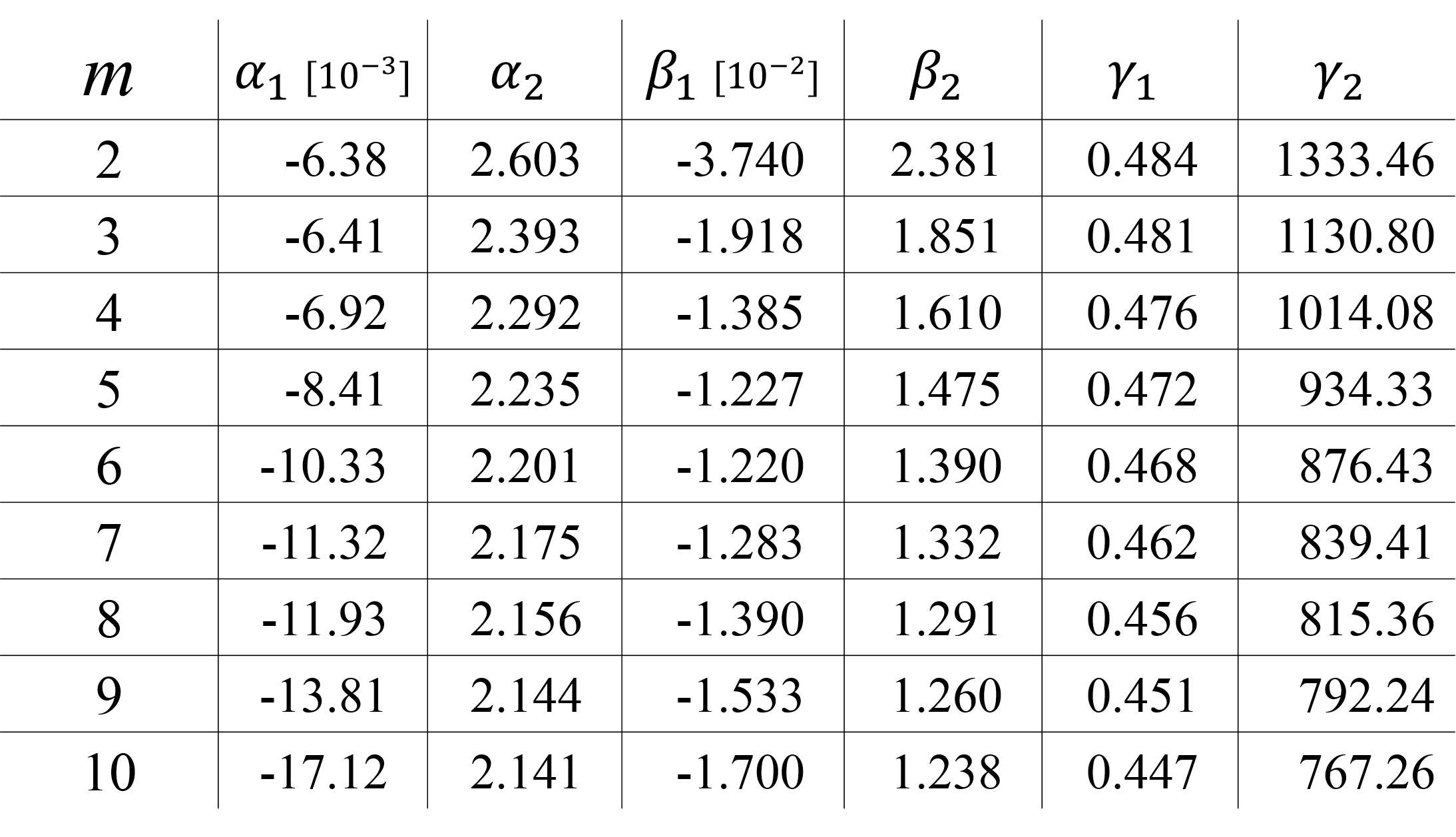}
	\caption{Exact values of $\alpha, \beta, \gamma$ with different $m$. One can use the given values to evaluate the ECE spectrum.}
	\label{table2}
\end{table}

\end{document}